\documentclass[preprint]{aastex}

\usepackage{natbib}
\begin{document}

\title{The Diffuse Source at the Center of LMC SNR 0509-67.5 is a Background Galaxy at z = 0.031}
\author{Ashley Pagnotta}
\affil{Department of Astrophysics, American Museum of Natural History, New York, NY 10024}
\author{Emma S. Walker}
\affil{Department of Physics, Yale University, New Haven, CT, 06520}
\and
\author{Bradley E. Schaefer}
\affil{Department of Physics and Astronomy, Louisiana State University, Baton Rouge, LA 70803}
\email{pagnotta@amnh.org}

\begin{abstract}
Type Ia supernovae (SNe Ia) are well-known for their use in the measurement of cosmological distances, but our continuing lack of concrete knowledge about their progenitor stars is both a matter of debate and a source of systematic error. In our attempts to answer this question, we presented unambiguous evidence that LMC SNR 0509-67.5, the remnant of an SN Ia that exploded in the Large Magellanic Cloud $400 \pm 50$ years ago, did not have any point sources (stars) near the site of the original supernova explosion, from which we concluded that this particular supernova must have had a progenitor system consisting of two white dwarfs \citep{schaefer2012a}. There is, however, evidence of nebulosity near the center of the remnant, which could have been left over detritus from the less massive WD, or could have been a background galaxy unrelated to the supernova explosion. We obtained long-slit spectra of the central nebulous region using GMOS on Gemini South to determine which of these two possibilities is correct. The spectra show H$\alpha$ emission at a redshift of $z=0.031$, which implies that the nebulosity in the center of LMC SNR 0509-67.5 is a background galaxy, unrelated to the supernova.
\end{abstract}


\section{The Type Ia Supernova Progenitor Problem}
\label{sec:intro}
Type Ia supernovae (SNe Ia) are well known standard(izable) candles used to measure cosmological distances, most famously in the discovery of the acceleration of the expansion of the universe \citep{riess1998a,perlmutter1999a}. Recent work has improved the statistical errors on the acceleration measurements to $\sim$6\%, however the systematics remain at $\sim$10\% \citep{howell2011a,sullivan2011a}. One way to reduce the systematics is to identify the progenitor system(s) of SNe Ia, allowing for more precise models of the physics and better calibration of the luminosity relations. For years we have hypothesized that the exploding star is a carbon-oxygen white dwarf (WD), which was recently confirmed by \citet{bloom2012a}, which must approach or exceed the Chandrasekhar mass limit \citep{chandrasekhar1931a}. The identity of the companion star that drives the WD over that limit has remained a hotly-debated mystery that has been noted as one of the top questions in astronomy \citep{blandford2010a}.

Many binary systems have been proposed as SN Ia progenitors, and they can loosely be divided into two main categories, the double-degenerates, consisting of two WDs \citep{tutukov1981a,van-kerkwijk2010a}, and the single-degenerates, consisting of one WD and a non-degenerate companion star, such as a main-sequence star or a red giant. We have seen very few WD-WD systems, as they are faint, but we have observed a variety of potential SD systems:  recurrent novae, symbiotic stars, supersoft X-ray sources, helium stars, spin-up/spin-down systems, and ``white widow" systems \citep{hachisu2001a,hachisu1999a,hachisu1999b,langer2000a,wang2009a,justham2011a,di-stefano2011a,wheeler2012b}. The ``currently favored model" is a matter that remains hotly debated, with strong proponents of many different solutions (e.g. \citealp{ruiz-lapuente2004a,gonzalez-hernandez2009a}, but see \citealp{kerzendorf2009a}; \citealp{edwards2012a}; \citealp{soderberg2012a}; \citealp{kushnir2013a}, \citealp{graur2013a}, and so forth).

\section{LMC SNR 0509-67.5}
\label{sec:snr0509}

The most definitive observational evidence so far comes from only one system, SNR 0509-67.5, which is the remnant of an SN Ia that exploded $400 \pm 50$ years ago in the Large Magellanic Cloud LMC) and was shown to be an over-luminous Type Ia of the SN1991T subclass \citep{hughes1995a,rest2005a,rest2008a,badenes2009a}. Taking into account the uncertainties on identifying the actual explosion site, the orbital velocity of all possible ex-companions, and any kick velocities imparted by the exploding WD, we used HST H$\alpha$ images from ACS to define a region near the center of LMC SNR 0509-67.5 in which we could reasonably expect to see any possible ex-companion star \citep{schaefer2012a}. Using HST WFC3 images in $B$, $V$, and $I$, we showed that there were no point-sources (with point-spread-functions consistent with stars) within this region down to a limiting magnitude of $V=26.9$ mag, which corresponds to an absolute magnitude of $M_V=+4.2$ mag at the accepted distance to the LMC. This means that there are no stars brighter than a K9 main-sequence star within the central region of LMC SNR 0509-67.5, allowing us to exclude all possible single degenerate models and conclude that this particular supernova must have had a double-degenerate progenitor system \citep{schaefer2012a}.

\begin{figure}
\centering
\epsscale{1.0}
\plotone{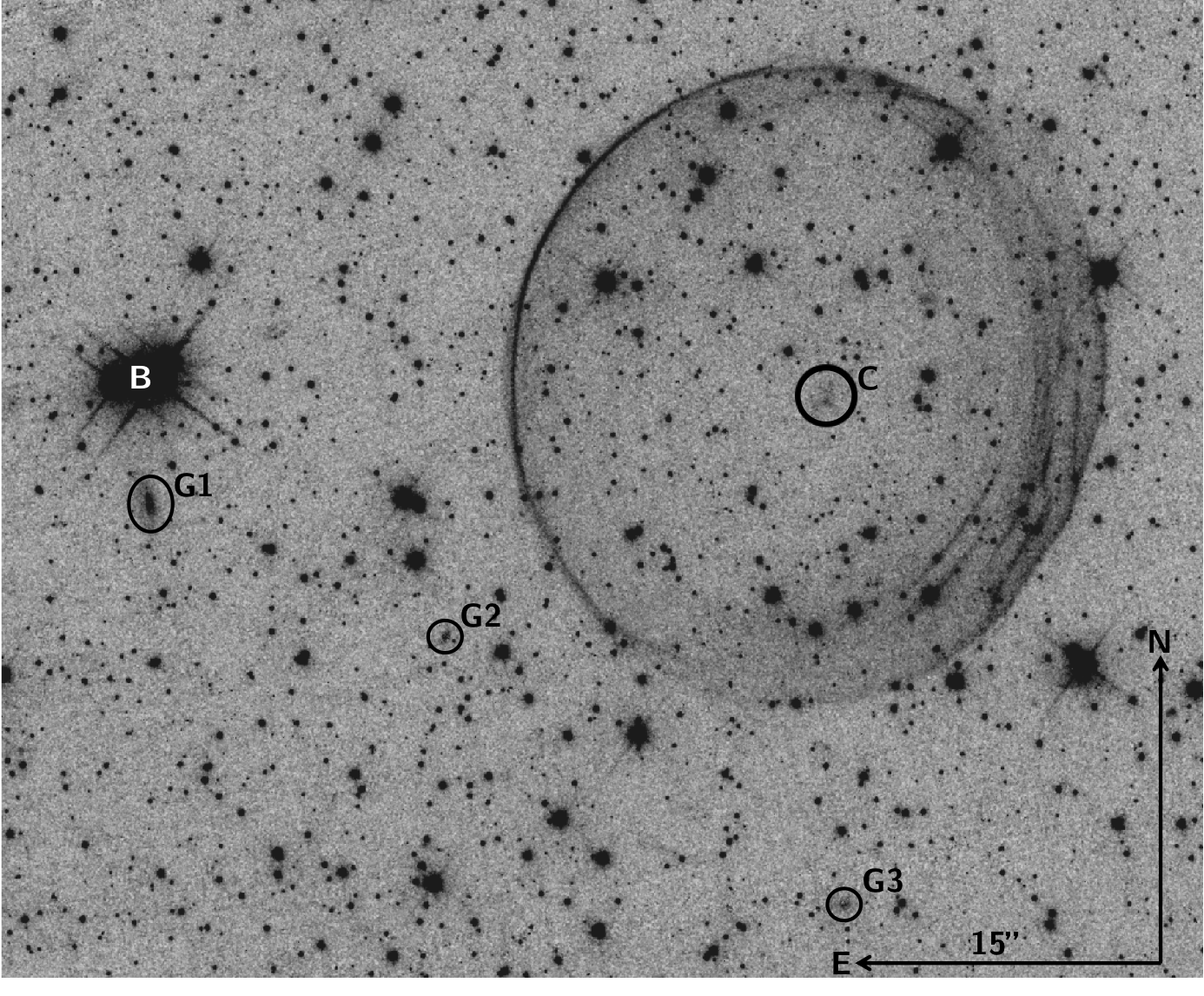}
\caption{This combined {\it BVI}+H$\alpha$ image shows the entirety of the LMC SNR 0509-67.5 remnant as well as the extreme 3$\sigma$ (99.73\% confidence) central region (marked C) within which any possible ex-companion star must be located. The nebulous smudge located within this region is clearly visible. It is not bright in the H$\alpha$ image, although it is very red in color, with $V=23.32 \pm 0.07$, $I=20.95 \pm 0.02$. Also marked in this image are the blind offset star used to obtain the GMOS spectrum (B) and three nebulous regions similar to the one in the center of the remnant that are background galaxies, for comparison (G1, G2, and G3).}
\label{fig:snr0509}
\end{figure}

Although we can  definitively state that there are no point sources visible in the HST observations of the central region of LMC SNR, there is a very obvious diffuse source, brightest in $I$-band ($V=23.32 \pm 0.07$, $I=20.95 \pm 0.02$), which can be seen in the combined $BVI$+H$\alpha$ image shown in Figure \ref{fig:snr0509}, contained within the central region, denoted C. As discussed in detail in \cite{schaefer2012a}, this nebulous ``smudge" cannot be a star, nor can it hide any stars behind it, thanks to the excellent resolution of HST. The most obvious origin for this object is that it is a background galaxy that just happens to be coincident with the central region of the SNR. A glance at the full HST image shows a number of similar irregular nebulous areas, mostly red, that also appear to be background galaxies, scattered throughout the region. Three of them are marked in Figure \ref{fig:snr0509}, as G1, G2, and G3. In the full 4500 arc-second squared field of view, there are five such red extended objects, and we therefore calculate a 0.9\% chance of such an object appearing within our central region. Although the nebulosity is not centered precisely at the location we calculate for the original explosion site, it is fully contained within the 3$\sigma$ error circle, the central region C in Figure \ref{fig:snr0509}, and was therefore worthy of further study to determine whether its location was purely coincidental alignment with a background galaxy or related to the supernova.

Although we usually describe the DD scenario in the very broad terms of ``two white dwarfs inspiraling and colliding, exceeding the Chandrasekhar mass limit (M$_\mathrm{Ch}$) and triggering the SN Ia explosion", the situation is, of course, not so simple, and at this time very difficult to model. The scenario proposed by \citet{piersanti2003a} describes a situation in which the heavier, primary WD tidally disrupts the lighter, secondary companion WD as they get closer. This process shreds the secondary and forms what is essentially an accretion disk composed of the material from that secondary WD, primarily carbon and oxygen. Material from the disk is accreted in the usual fashion until the primary WD exceeds M$_\mathrm{Ch}$ and explodes as an SN Ia. The unaccreted material from the secondary WD is massive enough that it is not dispersed by the actual Ia explosion, so it remains near the original explosion site, which in the case of LMC SNR 0509-67.5 would put it within the central region occupied by the extended source.

Here we report the results of our spectroscopic investigation of this central nebulosity to determine whether it is a background galaxy or the remnants of the shredded secondary WD. We would expect a galaxy to show a redshifted hydrogen line, while leftover WD material should show strong carbon and oxygen lines.

\section{Observations}
\label{sec:observations}

A spectrum was obtained using the GMOS instrument in long-slit mode with the R400 grism at a central wavelength of 6000{\AA} at Gemini-South\footnote{Program: GS-2011B-Q-30. PI: Schaefer} \citep{hook2004a} on 2011 November 5. The central source of the remnant was acquired by a blind offset from a nearby bright point source, which is marked as B in Figure 1. As such, the data were not obtained near parallactic angle. A slit of 2" width was used to maximize the light obtained from the faint source. The data were reduced and extracted using IRAF with standard and Gemini-specific routines. We present a spectrum which is a median of three exposures of 900s. Although the source is faint, we found the better signal-to-noise ratio obtained from taking the median made discerning spectral features easier than when co-adding the individual spectra to go deeper.  We use the median instead of the mean to reduce the effect of any cosmic rays.

The expected carbon and oxygen emission features depend on the physical conditions at the center of the remnant.  If the density has remained high, we would expect a low ionization state, and the [OI] 6300{\AA} line would be visible.  In contrast, if the density at the center is lower, the ionization state will be higher and we would see [OIII] 5007{\AA} and possibly CIII 4650{\AA} or CIV 4658{\AA}.  Upon examination of the optical spectrum, we detect none of these species.

In Figure 2, we show part of the spectrum of the the faint source at the center of the slit.  The grey region is the gap between the GMOS chips and we have set the data value to zero here.  The 1D spectrum shown below is the median of the pixel values between the two black lines shown on the 2D plot along with the derived wavelength solution.  In both the 1D and 2D plots we have rebinned the data from a $1\times1$ binning in the raw science frames to a $2\times2$ binning to increase the signal-to-noise.

In the region between the two black lines there is a flat continuum, as well as strong H$\alpha$ emission from the remnant (marked at $z=0$) detected at a level of $20\sigma$ above the continuum.  There is also a single extended emission line visible in the 2D image which you also see in the 1D spectrum detected at a significance of $6\sigma$.  We identify this line as H$\alpha$ at redshift $z=0.031$ and, along with evidence of a red, rather than blue, continuum emission (i.e., not hot material left from an explosion), we conclude that the source is indeed a background galaxy, rather than debris from a supernova explosion.  We do not confidently detect any other emission lines from the background galaxy, such as H$\beta$ or [OIII], but we also do not detect any emission lines of oxygen or carbon (permitted or forbidden) at $z=0$.  Although the lack of [OIII] is unusual, it is not altogether unsurprising, because [OIII] is often very faint or nearly non-existent in some types of galaxies, especially irregular ones. Examples of this can be seen in Figures 8 and 14 of \mbox{\cite{kennicutt1992a}}, especially in the spectrum of the NGC 3034 (M82). Since our spectrum of the nebulosity coincident with the center of LMC SNR 0509-67.5 has such low signal, even if there were [OIII] present, it could very easily be lost in the noise. Assuming a redshift of $z=0.031$ and measuring the approximate size of the H$\alpha$ emitting region from Figure 1 to be 2.14", this corresponds to a physical size of 1.3 kpc, assuming a cosmology of $(\Omega_M,\Omega_{\Lambda},H_0) = (0.72,0.28,72)$.

With only one line in the spectrum, our identification (H$\alpha$ at low redshift) is not fully constrained. The line is not anywhere near the wavelength we would expect for any line of a shredded WD disk at the distance of the LMC, and we have a non-detection of either carbon or oxygen species at $z=0$, so that hypothesis must be ruled out. The observed line, however, does match the prediction for a low-redshift background galaxy, and if it were any other common galactic spectrum line (other than H$\alpha$), it would result in an even higher redshift and still confirm the source as a background galaxy.

\begin{figure}
\centering
\epsscale{1.0}
\plotone{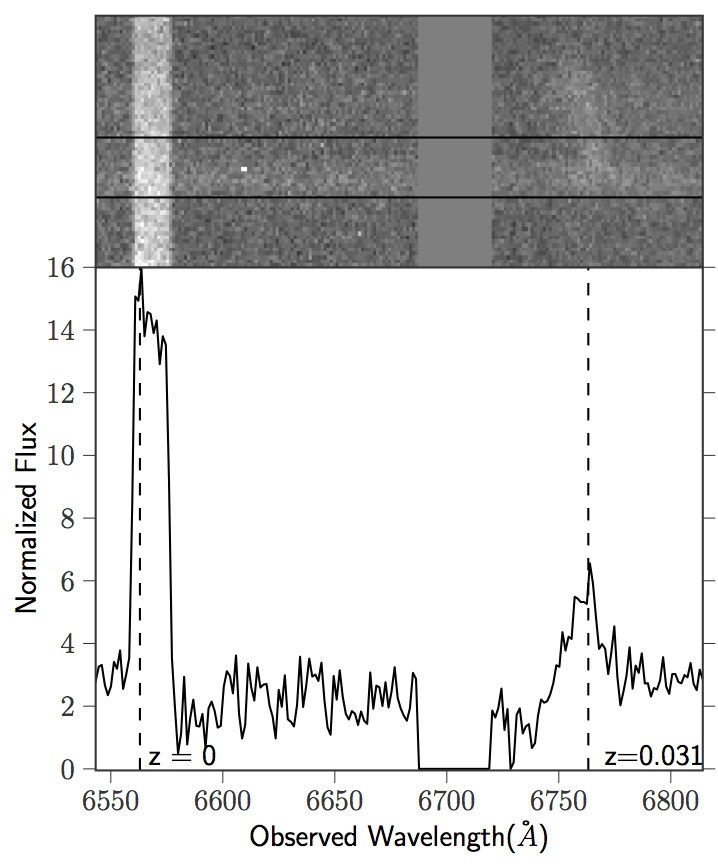}
\caption{Top: Part of the 2D spectrum of SNR0509-67.5.  The horizontal black lines have been added to highlight the continuum detection and to show the region used to generate the 1D spectrum below.  The grey block to the right of center is the gap between two CCDs in the GMOS detector.  This region has been set to have zero value. Bottom: The 1D spectrum of the source in the 2D image.  The vertical dashed lines show the wavelengths of H$\alpha$ at $z=0$ and $z=0.031$, which correspond to the edge of the remnant and the faint emission line from the central nebulosity, respectively.}
\label{fig:spectrum}
\end{figure}

\section{Conclusion}
\label{sec:conclusions}

The nebulosity seen in the central region of LMC SNR 0509-67.5 is due to a coincidentally located background galaxy. The spectrum is of low signal-to-noise, as is to be expected from a faint source and short integration time, but the single emission line from the central region of the remnant is confidently detected in the spectrum shown in Figure \ref{fig:spectrum}. We have identified this line as H$\alpha$ at a redshift of $z = 0.031$, as expected for a background galaxy. The alignment with the central region of the SNR is purely coincidental, and has no bearing on the identity of the progenitor system of the Type Ia supernova that created the remnant.

\acknowledgments
This research was supported by the National Science Foundation, the Louisiana State University Graduate School, and the Kathryn W. Davis Postdoctoral Scholar program. We would like to thank Sarah M. H{\"o}rst, Andrew S. Rivkin, Erin Lee Ryan, and all of the \#AcaDoWriMo participants. Additionally, we extend our gratitude to Maria M. Novilla.

%\bibliography{mybib}

\end{document}